\documentclass[a4paper,fleqn]{cas-dc}
\let\printorcid\relax % Remove ORCID footnote
\usepackage[authoryear,longnamesfirst]{natbib}
\usepackage{verbatim}
\usepackage{graphicx}
\usepackage{url}
\usepackage{import}
\usepackage{subcaption}
\usepackage{multirow}
\usepackage{color, colortbl}
\definecolor{Gray}{gray}{0.9}
\usepackage{color,soul}
\usepackage{tabularray}
\usepackage{breakurl}
\usepackage{nomencl}
\usepackage{framed}

\begin{document}
\let\WriteBookmarks\relax
\def\floatpagepagefraction{1}
\def\textpagefraction{.001}
\shorttitle\relax
\shortauthors{Aharon et~al.}  
\title [mode = title]{RoBERTa-Augmented Synthesis for Detecting Malicious API Requests}

\author[1,3]{Udi Aharon}
\cormark[1]
\author[5]{Revital Marbel}
\author[1,3]{Ran Dubin}
\author[1,3]{Amit Dvir}
\fnmark[1]
\author[2,3,4]{Chen Hajaj}

\cortext[1]{Corresponding author at: Department of Computer and Software Engineering, Ariel University, Golan Heights 1, 4077625, Ariel, Israel. E-mail addresses: udi.aharon@msmail.ariel.ac.il (U. Aharon), marbelr@hit.ac.il (R.Marbel) ,rand@ariel.ac.il (R. Dubin), amitdv@ariel.ac.il (A. Dvir), chenha@ariel.ac.il (C. Hajaj).}

\affiliation[1]{organization={Department of Computer and Software Engineering},
            addressline={Golan Heights 1}, 
            city={Ariel},
            postcode={4077625}, 
            country={Israel}}
            
\affiliation[2]{organization={Department of Industrial Engineering \& Management, Ariel University},
            addressline={Golan Heights 1}, 
            city={Ariel},
            postcode={4077625}, 
            country={Israel}}
            
\affiliation[3]{organization={Ariel Cyber Innovation Center, Ariel University},
            addressline={Golan Heights 1}, 
            city={Ariel},
            postcode={4077625}, 
            country={Israel}}

\affiliation[4]{organization={Data Science and Artificial Intelligence Research Center, Ariel University},
            addressline={Golan Heights 1}, 
            city={Ariel},
            postcode={4077625}, 
            country={Israel}}

\affiliation[5]{organization={Department of Computer Science, Holon Institute of Technology (HIT)},
            addressline={Golomb 52}, 
            city={Holon},
            postcode={4077625}, 
            country={Israel}}

\fntext[fn1]{IEEE Senior Member}         


\begin{abstract}
Web applications and APIs face constant threats from malicious actors seeking to exploit vulnerabilities for illicit gains. To defend against these threats, it is essential to have anomaly detection systems that can identify a variety of malicious behaviors. However, a significant challenge in this area is the limited availability of training data. Existing datasets often do not provide sufficient coverage of the diverse API structures, parameter formats, and usage patterns encountered in real-world scenarios. As a result, models trained on these datasets often struggle to generalize and may fail to detect less common or emerging attack vectors. To enhance detection accuracy and robustness, it is crucial to access larger and more representative datasets that capture the true variability of API traffic. To address this, we introduce a GAN-inspired learning framework that extends limited API traffic datasets through targeted, domain-aware synthesis. Drawing on techniques from Natural Language Processing (NLP), our approach leverages Transformer-based architectures, particularly RoBERTa, to enhance the contextual representation of API requests and generate realistic synthetic samples aligned with security-specific semantics. We evaluate our framework on two benchmark datasets, CSIC 2010 and ATRDF 2023, and compare it with a previous data augmentation technique to assess the importance of domain-specific synthesis. In addition, we apply our augmented data to various anomaly detection models to evaluate its impact on classification performance. Our method achieves up to a 4.94\% increase in F1 score on CSIC 2010 and up to 21.10\% on ATRDF 2023. The source codes of this work are available at~\url{https://github.com/ArielCyber/GAN-API}.

\end{abstract}

\begin{highlights}
\item Framework inspired by Generative Adversarial Networks augments web API traffic data.

\item RoBERTa-based language model enhances the semantics of synthetic web request samples.

\item Domain-specific synthesis improves anomaly detection on public HTTP traffic datasets.

\end{highlights}


\maketitle

\section{Introduction}

Application Programming Interfaces (APIs) are essential to modern software systems, facilitating structured communication and data exchange among distributed components and services~\cite{9723009,10704598}. The trend in software development is becoming increasingly modular, leveraging a diverse ecosystem of third-party services, modules, and tools~\cite{wijayarathna2019using,10612298,10.1145/2896587}. APIs serve as standardized access points to these components, allowing developers to integrate advanced functionalities, such as authentication and payment processing, without needing to understand their internal logic~\cite{GARRIGA201632}. This abstraction accelerates development cycles, encourages code reuse, and supports the creation of complex systems using modular building blocks~\cite{depa2025evolution}.

Among the various API paradigms, Representational State Transfer (REST) has gained popularity as a widely used architectural style for exposing and consuming services across heterogeneous environments~\cite{shyam2025performance,meshram2021evolution}. RESTful APIs are now integral to cloud platforms, mobile applications, and Internet of Things (IoT) ecosystems, offering lightweight, scalable interfaces that align with the HTTP protocol~\cite{mukherjee2024restful,CASADEI2025101548,pautasso2010restful}. Their simplicity and standardization make them ideal for enabling interoperability in distributed systems where flexibility and loose coupling are essential~\cite{meshram2021evolution}.

%RESTful APIs have become essential to modern web applications, serving as the primary means of interaction between the user interface and the server-side logic and orchestrating microservicess~\cite{10592777}. Web services designed on REST principles gain advantages from a uniform interface, stateless interactions, and a resource-oriented approach features that improve scalability, maintainability, and performance~\cite{pautasso2010restful}.

As the use of APIs has increased, so has the attack surface for adversaries seeking to exploit vulnerabilities~\cite{PAUL2024103871}. APIs often expose access points to users and systems, which can be targeted for malicious activities such as unauthorized access, data leaks, or service disruption~\cite{sun2021research,hussain2019current}. Common attack vectors that pose threats to API users include injection attacks, where unvalidated input lets attackers execute harmful commands or access data; Man-In-The-Middle (MitM) attacks, where communications between two parties are intercepted and altered without their knowledge; and Denial of Service (DoS) attacks, which bombard the API server with excessive requests, rendering it inaccessible to legitimate users~\cite{al2023man}. These types of attacks highlight the need for strong security measures, particularly for real-time anomaly detection and automated threat remediation, to protect the integrity, availability, and privacy of API interactions~\cite{idris2022web,YANG2025126982,AHARON2025104249}.

API security is still a developing research area, especially when compared to the broader field of web security, where machine learning and deep learning techniques have already been widely explored to detect threats~\cite{ABOLFATHI2024103673,alaoui2022deep,pașca2023overview}. A common challenge across many security applications, however, is the lack of open, labeled datasets~\cite{ZHOU2025104127,zhou2025webguard,dash2024deep}. Privacy concerns and corporate policies often prevent the release of web logs, and generating realistic synthetic data is challenging~\cite{guerra2022datasets}. As a result, much of the existing research relies on outdated or narrowly scoped datasets, limiting its ability to reflect the diversity and complexity of real-world environments accurately. This challenge is even greater with API datasets, as the structure and behavior of APIs can vary significantly across different systems. Addressing the vast array of methods, parameters, and routes—each with its context, expected input, and access patterns is a complex task.

Several methods have been introduced to address the challenge of limited datasets in deep learning. Few-shot learning (FSL) is specifically designed to enable models to learn effectively from small amounts of labeled data, making it particularly useful for tasks such as anomaly detection~\cite{yu2020intrusion,9016599,10082596}. On the other hand, Generative Adversarial Networks (GANs) are utilized to augment limited datasets by generating realistic synthetic samples before training~\cite{sun2020novel,saxena2021generative,chavdarova2018sgan,mohebbi2023games}. A GAN consists of two components: a generator, which produces synthetic data, and a discriminator, which assesses its authenticity. Through this adversarial process, the generator improves its ability to create increasingly realistic outputs. In addition to traditional GANs, several GAN-inspired~\cite{anidjar2024extending,xu2021gan,sundar2023gan} approaches have been proposed for more specialized use cases. These models enhance data diversity and control, address training instability, and reduce the computing costs of GANs for improved practicality.

Despite the progress made in prior research on synthetic data generation and few-shot learning for security applications, these methods have seen limited adaptation in the context of API traffic. While GANs offer a powerful mechanism for dataset augmentation, they are notoriously difficult to train, requiring extensive tuning, large volumes of initial data, and careful balancing between the generator and discriminator. GAN-inspired approaches, which simplify or reimagine the adversarial process, present a more practical alternative and are often designed with domain-specific knowledge in mind. However, API traffic introduces unique challenges that require specialized handling, including structured request formats, diverse parameter types, contextual dependencies, and domain-specific vocabularies.

In this context, we propose a novel GAN-inspired learning framework for API attack detection that leverages Robustly Optimized BERT Approach (RoBERTa)~\cite{liu2019roberta}, a Transformer-based architecture, to improve the contextual representation of API requests and generate synthetic samples aligned with security-specific semantics. By treating structured API requests as language-like sequences, our model learns rich contextual representations that capture both syntax and intent. This representation is then used to generate realistic synthetic API samples through a lightweight generator module that mimics the data expansion role of traditional GANs, without requiring the full adversarial training pipeline. This paper's contributions are itemized as follows: 

\begin{enumerate}[\textbullet]
\item We propose a GAN-inspired framework that retains core GAN principles in a lightweight design, enabling the generation of high-quality, diverse API requests from a limited dataset.

\item Our approach leverages RoBERTa as both a generator guided by API-specific constraints and a discriminator trained in a self-supervised manner using Pseudo Labeling (PL) and Out Of Distribution (OOD) scoring to distinguish between normal and anomalous API requests.

\item We introduce a domain-specific token masking strategy based on semantic distance and statistical filtering, ensuring that generated API samples remain structurally valid and contextually relevant to the API domain.

\item We evaluate our framework on two benchmark datasets, CSIC 2010 and ATRDF 2023, and compare it with a previous data augmentation technique to assess the importance of domain-specific synthesis.

\item We apply our augmented data to various anomaly detection models to evaluate its impact on classification performance.

\end{enumerate}

The remainder of this paper is organized as follows. Section~\ref{sec:related_work} reviews prior work related to anomaly detection, few-shot learning, and deep learning techniques for data augmentation. Section~\ref{sec:architecture} details the design and components of the proposed framework. The datasets used in our experiments are introduced in Section~\ref{sec:datasets}, followed by an explanation of the evaluation metrics in Section~\ref{sec:evaluation_metrics}. Section~\ref{sec:results} presents our empirical findings and highlights the framework’s performance. Finally, Section~\ref{sec:conclusions} summarizes the key contributions, discusses current limitations, and outlines directions for future research. For reference, Table~\ref{tab:abbreviations} lists the key abbreviations used throughout the paper.

\begin{table}[]
\resizebox{\columnwidth}{!}{%
\begin{tabular}{ll}
        \hline
        \textbf{Abbreviation} & \textbf{Definition} \\
        \hline
        AE & Autoencoder \\
        API & Application Programming Interface \\
        BBPE & Byte-Level Byte Pair Encoding \\
        BERT & Bidirectional Encoder Representations from Transformers \\
        FSL & Few-shot Learning \\
        GAN & Generative Adversarial Network \\
        MLM & Masked Language Modeling \\
        MCC & Matthews Correlation Coefficient \\
        NLP & Natural Language Processing \\
        OOD & Out Of Distribution \\
        RoBERTa & Optimized BERT Approach \\
        SOTA & State-of-the-Art \\
        SQLi & SQL Injection \\
        XSS & Cross-Site Scripting \\
\end{tabular}%
}
\caption{List of Abbreviations.}
\label{tab:abbreviations}
\end{table}

\section{Related Work}
\label{sec:related_work}
In this section, we provide a comprehensive review of relevant research across three key areas that form the foundation of our proposed framework. First, we examine existing web security and anomaly detection work, which provides the broader context for understanding API threats. Since a large portion of modern APIs, particularly RESTful APIs, communicate over HTTP, many techniques developed for web traffic analysis, are directly applicable to API traffic. These challenges underscore the need for tailored detection approaches that can adapt to the structural and behavioral nuances of API communications. Second, we review FSL techniques, which aim to train models effectively with limited labeled data. Third, we discuss deep learning techniques for data augmentation.

\subsection{Web Security and Anomaly Detection}
Extensive research has been conducted on detecting malicious behavior in web traffic, including attacks such as SQL Injection (SQLi), and Cross-Site Scripting (XSS). While early approaches often relied on rule-based or statistical models, more recent work has adopted machine learning and NLP techniques to analyze structured web requests and detect anomalies.

Several early works applied classical machine learning techniques combined with statistical features. Yao et al.~\cite{pan2019} introduced a synthetic dataset mimicking internal web application traffic and applied n-gram analysis with supervised models such as Naive Bayes, Random Forest, and Support Vector Machine (SVM), as well as Autoencoders (AEs) for unsupervised detection. Zhang et al.~\cite{zhang2019} focused on identifying input validation and sanitization in PHP files using both traditional classifiers and CNNs. Xie et al.~\cite{xie2019} proposed a method for detecting SQLi attacks using Word2Vec embeddings and an Elastic-Pooling Convolutional Neural Network (CNN). Ashlam et al.~\cite{ashlam2022} presented a lightweight model using stop-word removal, CountVectorizer, and various classifiers to detect SQLi attempts in HTTP traffic.

Building on this foundation, more recent studies have employed deep learning and word embedding techniques for richer semantic representation. Li et al.~\cite{li2019} proposed a web application attack detection model using attention and gated convolutional networks, enhanced with Structure-Level Segmentation and word embeddings. Chen et al.~\cite{chen2021} used Word2Vec to represent serialized input formats such as JSON and PHP, followed by a CNN and Multilayer Perceptron (MLP) architecture for classification. Gniewkowski et al.~\cite{gniewkowski2023sec2vec} introduced a semi-supervised language representation approach for HTTP request embeddings using Bag-of-Words (BoW), fastText, and RoBERTa, which were then applied for anomaly classification.

More recent advancements have focused on applying Transformer-based models for improved representation and sequence understanding, particularly in URL and request analysis. Maneriker et al.~\cite{maneriker2021} proposed URLTran, a Transformer-based model tailored for phishing URL detection. Sanwal and Ozcan~\cite{sanwal2021} introduced a hybrid model based on CharacterBERT, a variant of Bidirectional Encoder Representations from Transformers (BERT) designed to generate word-level contextual embeddings. They used it to capture non-linear embeddings from URLs for phishing detection.

\subsection{Few-Shot Learning}
The application of deep learning to cybersecurity tasks has drawn significant attention to a persistent limitation - insufficient labeled data, especially for rare or emerging cyberattacks. FSL has emerged as a solution to this problem. It seeks to build accurate models using only a small number of samples, effectively simulating realistic cybersecurity conditions where full annotations are impractical or unavailable.

One line of work focuses on improving few-shot performance through discriminative representation learning. Iliyasu et al.~\cite{iliyasu2022few} propose a supervised AE trained jointly on reconstruction and classification loss to learn structured representations. These features are then passed to a lightweight linear classifier for few-shot inference. Similarly,~\cite{nguyen2023few} introduces Discriminative Representation-based AE, which incorporates a margin-based constraint in the latent space to encourage separation between normal and anomalous traffic. The learned embeddings are used in a meta-learning setting to classify new attack types using only a few labeled examples.

Another direction leverages meta-learning to enable rapid adaptation to unseen tasks. Meta-learning, often described as "learning to learn", involves training a model across a distribution of tasks so that it can quickly adapt to new ones with minimal data~\cite{tian2022meta}. Yang et al.~\cite{yang2023few} propose MetaMRE, which combines multi-task learning with flow-level representation and a clustering-based prototype refinement mechanism for classifying encrypted traffic under limited supervision. Wang et al.~\cite{wang2024few} present an online meta-learning framework for intrusion detection in IoT environments. Their system incorporates an anti-forgetting mechanism to support continual learning, allowing the model to adapt to new attacks while preserving knowledge from earlier tasks.

Other studies adopt application-specific approaches to implement few-shot classification in structured domains. A study on XSS detection~\cite{pan2024few} formulates payload category recognition as a few-shot task and applies a neural classifier trained on limited examples. Yi et al.~\cite{yi2024nfhp} encodes network traffic as two-dimensional holographic images and uses a ResNet-based architecture trained through few-shot episodes. Yan et al.~\cite{yan2023gde} employ a hybrid attention mechanism and lightweight design to perform variable intrusion detection using small support sets.

While these studies demonstrate the potential of FSL for security applications, many rely on predefined support sets, well-separated feature spaces, or consistent task structures. In practice, these assumptions often break down in API traffic, where requests may vary widely in format, semantics, and behavior. 

%Semeniuta et al.~\cite{semeniuta2017hybrid} proposed a hybrid VAE architecture combining convolutional layers with recurrent decoders to facilitate optimization and improve long-sequence generation. 
%\subsection{GANs and GAN-Inspired Techniques for Data Augmentation}
\subsection{Deep Learning Techniques for Data Augmentation}
Deep learning techniques have become widely used for data augmentation, especially in domains with limited or imbalanced labeled data~\cite{iqbal2022survey}. A common approach involves Variational Autoencoders (VAEs)~\cite{kingma2019introduction}, which enable unsupervised text generation by learning latent representations of input sequences, making it possible to generate coherent and structured text samples. Early models such as Bowman et al.~\cite{bowman2015generating} introduced VAE architectures using Recurrent Neural Network (RNN)-based encoders and decoders to generate sentences from continuous latent spaces. However, these models exhibited posterior collapse, where powerful decoders ignore the latent variables, and required large-scale datasets to achieve effective performance. Several approaches have been proposed to address these limitations. Yang et al.~\cite{yang2017improved} explored dilated CNNs as decoders, balancing contextual capacity to better leverage latent information, though reliance on data scale remained a drawback. Hu et al.~\cite{hu2020variational} extended RNN-based VAEs by integrating variational inference into a language model, showing improvements in generating coherent sentences. Recent work~\cite{makelov2024evaluating} examined sparse AEs to control attributes in open-ended text generation. 

Another widely used generative framework is GANs, which consist of a generator that learns to produce synthetic data and a discriminator that evaluates how closely these samples resemble real data. While GANs have achieved notable success in image generation, their application to text has been more limited due to the discrete nature of language, which complicates gradient-based training~\cite{iqbal2022survey}. Early work by Li et al.~\cite{li2017adversarial} explored adversarial learning for neural dialogue generation using reinforcement learning (RL) to address exposure bias in maximum likelihood estimation. Jiao and Ren~\cite{jiao2021wrgan} incorporated Wasserstein loss to improve output quality. However, they observed that model performance is highly sensitive to the generator–discriminator update ratio, with improper tuning leading to instability and overfitting during training. Diao et al.~\cite{diao2021tilgan} introduced a Transformer-based GAN that implicitly models latent space distributions, achieving more fluent and diverse text. In the context of network security, Ding et al.~\cite{ding2022imbalanced} proposed a hybrid approach combining K-Nearest Neighbors (KNN) for undersampling with a GAN variant to oversample minority-class intrusion data. Similarly, Chapaneri and Shah~\cite{chapaneri2022enhanced} presented a GAN-based method to augment malicious network traffic, leveraging Wasserstein loss to improve training stability. Sahal et al.~\cite{sahal2024generating} applied GANs for synthetic text generation in bacterial sigma factor prediction but noted that effective training requires large volumes of data and incurs significant computational and time costs.

To overcome challenges such as instability, heavy computational demands, and the reliance on large datasets, several studies have proposed GAN-inspired techniques that simplify the training process while preserving the core adversarial principles of GANs. Anidjar et al.~\cite{anidjar2024extending} introduced a GAN-like self-supervision framework designed to extend limited SMS spam datasets. Their approach employs a generator-discriminator mechanism to produce synthetic samples that preserve distributional properties relevant to spam detection. In a different domain, Xu and Karamouzas~\cite{xu2021gan} proposed a GAN-like imitation learning method for interactive character control in physics-based animation. Their model replaces the explicit adversarial game with an ensemble of discriminators and a policy trained via reinforcement learning. Sundar et al.~\cite{sundar2023gan} applied GAN-inspired training to enhance security in federated learning systems. They designed a defense mechanism against backdoor attacks by simulating poisoned data distributions and training the system to recognize and reject them.

\section{Framework}
\label{sec:architecture}

In our work, we use Transformers, a deep learning architecture originally introduced for sequence modeling tasks~\cite{vaswani2017attention}. Transformers are well-suited for self-supervised learning and rely on self-attention mechanisms, which allow the model to weigh the importance of each token in relation to others in the sequence. This makes it possible to capture both semantic meaning and syntactic structure within the input. By using multiple attention heads, the model can learn different types of contextual relationships in parallel, resulting in more expressive and nuanced representations~\cite{islam2024comprehensive}. The layered design of Transformers further enables the model to capture higher-level patterns across longer spans of text, which is particularly effective for modeling structured API requests.

Transformer architecture is the backbone of various state-of-the-art (SOTA) language models, including BERT and RoBERTa~\cite{wolf2020transformers}. RoBERTa builds on the foundational BERT model, with some key updates in the way it’s trained to enhance its performance in language comprehension tasks. One of those updates is the use of a Byte-Level Byte Pair Encoding (BBPE) tokenizer. This tokenizer breaks the input text into byte-sized chunks, each carrying meaningful semantic information~\cite{wang2020neural}.

MLM is a training technique that encourages the model to learn representations that capture the semantics of language~\cite{liu2019roberta}. In MLM, some words in a sentence are randomly replaced with a special token, $<MASK>$, which is specifically used in the RoBERTa model. The goal is for the model to predict the original word based on the context provided by the remaining, unmasked words. Among the various NLP tasks supported by MLM, the fill-mask task allows a language model to infer and predict the masked token in a sentence using the surrounding context~\cite{kesgin2023iterative}.

\begin{figure*}[!htb]
\centering
\includegraphics[width=7in]{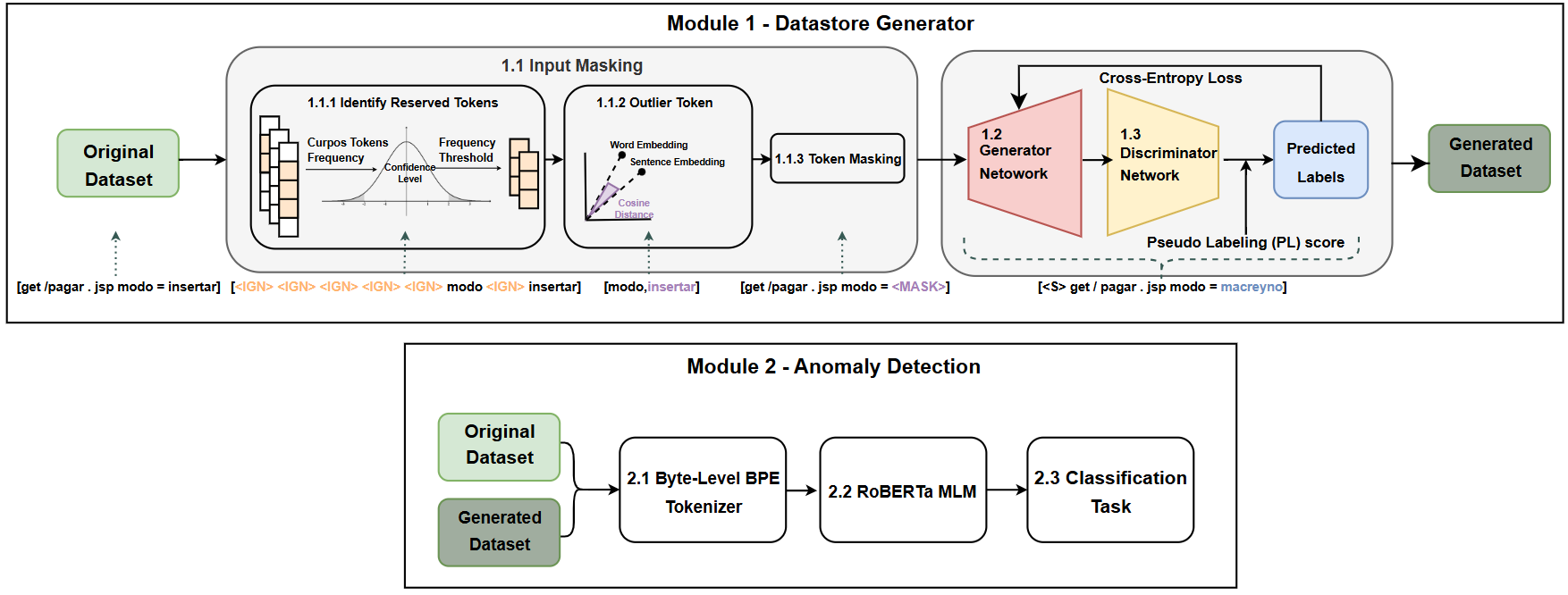}
\caption{Architecture of the proposed system. The system consists of two main modules: Module 1 – Datastore Generator, and Module 2 – Anomaly Detection. In Module 1, the original API request dataset is processed through a masking pipeline that identifies reserved tokens (marked as "<IGN>") based on frequency thresholds and excludes them from substitution. The remaining tokens are evaluated for semantic fit using cosine similarity between word and sentence-level embeddings, and the outlier token is replaced with a "<MASK>" token. In the example shown, "insertar" is identified as the outlier and replaced, resulting in the masked input "get /pagar.jsp modo=<MASK>". A generator network predicts suitable replacements, and a discriminator network assigns a PL confidence score to the generated variant. This results in a synthetic request, such as "get /pagar.jsp modo=macreyno", which is included in the augmented dataset if the discriminator assigns it a high-confidence score. Module 2 takes both the original and generated datasets, tokenizes them using a byte-level BPE tokenizer (2.1), processes them through a RoBERTa-based masked language model (2.2), and finally applies a classifier (2.3) to detect anomalies in API requests.}
\label{fig:architecture}
\end{figure*}

The proposed framework (depicted in Figure~\ref{fig:architecture}) comprises two modules: a datastore generator designed to augment sparse datasets by synthesizing realistic API requests, and a downstream anomaly detection module trained on the extended dataset that includes both original and generated samples. The first module leverages a RoBERTa-based architecture to generate realistic API requests from limited data, using token masking guided by structural and semantic constraints. Once the synthetic data is validated, it is used to train the second module, an anomaly detection model capable of classifying API requests as benign or malicious.

\subsection{Datastore Generator}
The datastore generator, as depicted in Figure~\ref{fig:architecture} in step 1, is responsible for synthesizing realistic API requests from a limited input dataset. It serves as the foundation for expanding sparse data environments by learning structural patterns and semantic dependencies within API requests. The process consists of three sequential stages: Input Masking, Generator Network, and Discriminator Network. First, the input masking step identifies tokens suitable for replacement based on frequency and semantic distance criteria. Then, the generator predicts suitable substitutes for the masked tokens. Finally, the discriminator assigns confidence scores, filtering the generated samples before inclusion in the extended dataset.

\subsubsection{Input Masking}
\label{sec:input_masking}
The input masking step, shown in Figure~\ref{fig:architecture} as step 1.1, involves masking one element within a sequence of input entities from an API request and training the model to predict the masked entity. An API request may contain a mix of ASCII characters, numbers, and words, among other components. The requests are first processed by splitting them into identifiable entities using ASCII punctuation and symbols, such as “/”. For each set of extracted entities, we identify the outlier token, the token that is least similar to the overall context of the request, based on cosine similarity, and mark it as a candidate for replacement. The detailed procedure is described in the following section.

\paragraph{Identify Reserved Tokens}
API functionality operates within a constrained structural language framework, which limits the range of characters and tokens it can effectively process. To address this, we identify reserved tokens that should be excluded during the generative process. These are marked using a new special token, $<IGN>$. As shown in Figure~\ref{fig:architecture}, step 1.1.1, we generate a token frequency distribution based on the training dataset and compute the $z$-score to establish a frequency threshold, a technique commonly used in outlier detection tasks~\cite{rousseeuw2011robust}. There are several possible ways to define a threshold, including rule-based cutoffs, empirical tuning, or iterative adjustment of $z$-score values to find an optimal standard deviation~\cite{zhou2021optimal}. In our work, we adopt a statistically grounded method by deriving the $z$-score directly from a predefined confidence level~\cite{simundic2008confidence}. We use a high confidence level of 99.99\% to minimize the likelihood of including items within the dataset with frequencies that could be attributed to random fluctuations or noise. Let $X$ be the token frequency, $C$ be the confidence level, and $Z'$ be the $z$-score. The threshold $T$ is defined as: $T = Mean(X) + (Z' * Std(X))$. By identifying the top frequent tokens through this method, we can identify tokens representing fundamental components of the HTTP protocol, including methods and URL components.

\paragraph{Outlier Token}
To identify the outlier token, as shown in Figure~\ref{fig:architecture}, step 1.1.2, we use the RoBERTa encoder to extract word embeddings for each token in the request and compute a sentence embedding representing the full API request. We then calculate the cosine similarity between each word embedding and the sentence embedding to determine which token is least aligned with the overall context~\cite{bear2023fine}. The outlier token is defined as the one with the lowest cosine similarity to the sentence embedding, excluding any reserved tokens as described earlier. Formally, for an API request $R$ of a size $n \in N$ tokens, the method seeks a token $t_i$, where $1 \leq i \leq n$, such that $cos-sim(R, t_i)$ is minimized.

\paragraph{Token Masking}
Once the outlier token within an API request has been identified, we replace it with the $<MASK>$ token and feed the modified request into RoBERTa to predict the masked word, as shown in Figure~\ref{fig:architecture}, step 1.1.3.

\subsubsection{Generator}
The proposed generator, as shown in Figure~\ref{fig:architecture}, step 1.2, is based on a MLM built on the RoBERTa architecture. It combines a RoBERTa encoder, a neural network component that processes an input sequence and generates contextual representations for each token based on its relationships with other tokens~\cite{liu2019roberta}, with an additional linear layer to produce a probability distribution over the vocabulary for each masked token. The generator uses the final hidden state corresponding to the $<MASK>$ token and uses the Softmax function to compute token probabilities across the vocabulary, following standard language modeling practices. The loss function used to train the generator is Cross-Entropy loss~\cite{ho2019real}, which is well-suited for masked word prediction.

\subsubsection{Discriminator}
The discriminator, as depicted in Figure~\ref{fig:architecture}, step 1.3, is a self-supervised deep neural network with a sentence-level attention head, designed to distinguish between normal API requests and anomalous ones generated by the model. Built on a RoBERTa encoder, it transforms token sequences into contextualized word embeddings. A linear layer is applied on top of the encoder to produce a probability distribution over output classes. The model uses the special $<CLS>$ token from the encoder’s final layer to generate sentence-level predictions, classifying requests as either normal or anomalous. To support this classification, we introduce an OOD score, derived from the probability vector produced by the linear layer. While OOD detection typically refers to identifying inputs that diverge from the training distribution~\cite{ren2019likelihood}, in our case, we use the OOD score as an internal signal to assess the confidence of the discriminator’s prediction. This enables a meaningful comparison between original API requests and generated API requests, allowing for evaluation of the impact of token substitution on semantic integrity. Since the generated requests are unlabeled, we infer their class using a confidence-based approach. We adopt the PL score~\cite{manolache2021date}, which assumes that requests with higher uncertainty, reflected in a more uniform probability distribution, are more likely to be anomalous. Rather than computing this score at the token level, we apply it at the request level by evaluating the sentence-level probability vector and applying a threshold to determine the classification outcome.

\subsection{Anomaly Detection}
Once the generated dataset is available, we proceed to the second module, as depicted in Figure~\ref{fig:architecture}, step 2, which involves training the anomaly detection model. We begin by merging the generated dataset with the original dataset. The requests are then processed similarly to the approach described in Section~\ref{sec:input_masking}. In step 2.1 of Figure~\ref{fig:architecture}, we train a BBPE tokenizer from scratch on the unified dataset. This approach helps produce a reliable tokenizer while reducing the risk of introducing bias into the anomaly detection process. Such bias may occur if the tokenizer generates an excessive number of short or fragmented tokens, which can artificially inflate the perceived irregularity of otherwise normal requests~\cite{song2023selfseg}. For instance, the path "/api/user/1" may be tokenized as ["/", "a", "p", "i", "/", "u", "s", "e", "r", "/", "1"] rather than as larger, semantically meaningful units, leading the model to misinterpret normal input as anomalous. Next, in step 2.2, we train a RoBERTa-based MLM from scratch using only training data corresponding to normal API requests. This allows the model to learn patterns and structural features that are characteristic of normal traffic, which is essential for effective anomaly detection. Subsequently, in Step 2.3 (Figure~\ref{fig:architecture}), we train a Random Forest classifier to perform the final anomaly detection task. Random Forest is an ensemble learning method that constructs multiple decision trees during training and outputs the majority vote of their predictions. As shown by Kirasich et al.~\cite{kirasich2018random}, Random Forest maintains high true positive rates even when a large portion of the input features are irrelevant or weakly informative. This aligns with our setting, where tokenized API requests may contain sparse or noisy elements. Additionally, the method effectively handles non-linear relationships between features and output classes, enabling it to capture the complex decision boundaries that often characterize malicious behavior.

\section{Experimental Design}
\label{sec:datasets}

We conducted our experiments primarily on two public datasets, namely CSIC 2010, and ATRDF 2023. The statistical information from our datasets is as shown in Table~\ref{tab:datasets}

\renewcommand{\arraystretch}{1.2}
\begin{table}[!ht]
\centering
\scalebox{0.9}{
\begin{tabular}{lllll}
\hline
\multicolumn{1}{c}{\multirow{2}{*}{Dataset}} & \multicolumn{2}{c}{Train} & \multicolumn{2}{c}{Test} \\ \cline{2-5} 
\multicolumn{1}{c}{} & \multicolumn{1}{l}{Normal} & Abnormal & \multicolumn{1}{l}{Normal} & Abnormal \\ \hline
ATRDF 2023 & \multicolumn{1}{l}{43,200} & {54,600} & \multicolumn{1}{l}{10,800} & {23,400} \\ \hline
CSIC 2010 & \multicolumn{1}{l}{50,400} & {17,545} & \multicolumn{1}{l}{21,600} & {7,520} \\ \hline
\end{tabular}
}
\caption{Dataset splits used in our experiments. Each dataset was divided into 70\% training and 30\% testing portions. The training split was used both for sample generation and for training the classifier. The RoBERTa model was trained exclusively on the normal portion of the training set.}
\label{tab:datasets}
\end{table}

\textbf{CSIC 2010}~\cite{gimenez2010http}. It is a benchmark dataset developed by the Institute of Information Security of the Spanish National Research Council (CSIC). It simulates HTTP traffic for an e-commerce web application, where users interact with features such as product searches, shopping carts, and registration forms. The dataset includes over 72,000 web requests, comprising 36,000 normal requests and more than 25,000 attack instances. These attack samples span a wide range of categories, including SQLi, buffer overflows, information gathering, file disclosure, Carriage Return Line Feed (CRLF) injection, XSS, server-side inclusion, and parameter tampering. The dataset has been widely used in previous studies~\cite{emanet2023ensemble,pardomuan2023server} to evaluate web application firewalls and intrusion detection systems. 

\textbf{ATRDF 2023}~\cite{atrdf2023lavian}. It is a comprehensive dataset of API traffic, consisting of both normal and malicious request-response pairs. It reflects modern API formats and includes a wide variety of payload-based attacks, such as Directory Traversal (DT), Cookie Injection (CI), LOG4J, Remote Code Execution (RCE), Log Forging (LF), SQLi, and XSS, all labeled accordingly. The dataset is structured into four distinct groups and comprises approximately 54,000 normal pairs and 78,000 abnormal pairs. It has been adopted in recent research~\cite{AHARON2025104249} for studying API-level threat detection and classification.

\begin{figure}[ht]
  \centering
  \includegraphics[width=1.5in]{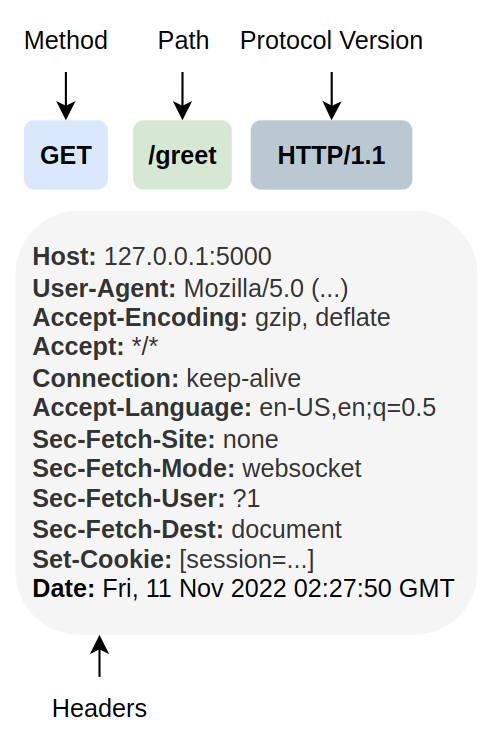}
  \caption{Example of a structured HTTP API request extracted from the ATRDF 2023 dataset, showing the method indicating the operation performed, the path identifying the requested resource, and the headers providing metadata such as content type, client identity, and session details.}
  \label{fig:api_request_example}
\end{figure}

Although CSIC 2010 represents HTTP traffic and ATRDF 2023 is explicitly API-based, we treat both datasets as HTTP-based API traffic, given that modern APIs rely on the same underlying protocol and structural patterns~\cite{niswar2024performance}. Figure~\ref{fig:api_request_example} provides an example of an API request, highlighting components such as the HTTP method, resource path, header fields, and request body. Table~\ref{fig:attack_examples} presents sample payloads for several common attack types, such as SQLi, XSS, and DT, as found in the ATRDF 2023 dataset.

\renewcommand{\arraystretch}{1.2}
\begin{table*}[!ht]
\centering
\caption{
Attack payload examples from the ATRDF 2023 dataset, categorized by type and vector, with corresponding descriptions.
}
\label{fig:attack_examples}
\resizebox{\textwidth}{!}{%
\begin{tabular}{
    >{\raggedright\arraybackslash}p{2.7cm} |
    >{\raggedright\arraybackslash}p{6cm} |
    >{\raggedright\arraybackslash}p{5cm} |
    >{\raggedright\arraybackslash}p{6cm}
}
\hline
\textbf{Attack Type} & \textbf{Example} & \textbf{Payload} & \textbf{Description} \\ \hline

Directory Traversal &
GET /static/download\_txt/
../../windows.ini.txt &
../../windows.ini.txt &
Attempts to read sensitive files by escaping the web root through a URL-based path traversal payload. \\ \hline

Cookie Injection &
Cookie: username=gASVyQAAAAAAAACMCG... &
username=...system("powershell echo 'hello world'") &
Encodes a malicious command in a cookie header field, allowing potential code execution when parsed on the server. \\ \hline

LOG4J &
Sec-Fetch-Site: none\$\{jndi:ldap2domeinold.ru:1223
/Payload\} &
\$\{jndi:ldap2domeinold.ru:1223
/Payload\} &
Targets header and input fields to exploit Java-based loggers through JNDI remote lookup using a crafted payload. \\ \hline

Remote Code Execution &
GET /greet/\{\{get\_flashed\_messages.
\_\_globals\_\_...system\%7D\%7D\}\} &
\{\{...\_\_import('os').system...\}\} &
Injects Jinja2 templates in the URL path to trigger server-side code execution via template rendering. \\ \hline

Log Forging &
GET /orders/check/exists?val=10403
\%0ASIGNOUT:\%20user\%20Shirley &
val=10403\textbackslash nSIGNOUT: user Shirley &
Uses header or query fields to inject newline characters, forging log entries and potentially misleading audit trails. \\ \hline

SQL Injection &
GET /orders/get/country?
country=';SELECT\%20*\%20FROM
\%20shippers\%20-- &
';SELECT * FROM shippers -- &
Appends raw SQL code into a query parameter to manipulate backend queries and extract database records. \\

\hline
\end{tabular}}
\end{table*}

\section{Evaluation Metrics}
\label{sec:evaluation_metrics}

To evaluate the quality of data augmentation in the generative stage, we use three complementary text generation evaluation metrics: BiLingual Evaluation Understudy (BLEU)~\cite{papineni2002bleu}, BERTScore~\cite{zhang2019bertscore}, and MoverScore~\cite{zhao2019moverscore}. BLEU measures surface-level similarity by computing n-gram overlap between generated and reference sequences, making it effective for capturing exact lexical matches. BERTScore computes semantic similarity using contextualized token embeddings from pretrained language models, making it well-suited for handling paraphrasing and variations in word order. MoverScore extends this approach by combining contextual embeddings with Earth Mover’s Distance~\cite{rubner2000earth}, which quantifies the minimum semantic effort required to transform one sequence into another and thus measures how closely the meaning of the generated text aligns with the reference.

For the anomaly detection task, we use four classification metrics: Precision, Recall, F1 Score, and Matthews Correlation Coefficient (MCC). These are defined as:

\par\textbf{Precision.} Quantifies the number of correctly predicted positive instances out of all predicted positives.
\[
\text{Precision} = \frac{TP}{TP + FP}
\]

\par\textbf{Recall.} Measures the proportion of actual positives correctly identified by the model.
\[
\text{Recall} = \frac{TP}{TP + FN}
\]

\par\textbf{F1 Score.} Represents the harmonic mean of Precision and Recall.
\[
\text{F1 Score} = \frac{2 \cdot \text{Precision} \cdot \text{Recall}}{\text{Precision} + \text{Recall}}
\]

\par\textbf{MCC.} Provides a balanced evaluation of classification performance, even with class imbalance, using all four elements of the confusion matrix.
\[
\text{MCC} = \frac{TP \cdot TN - FP \cdot FN}
{\sqrt{(TP + FP)(TP + FN)(TN + FP)(TN + FN)}}
\]

In these formulas, $TP$, $TN$, $FP$, and $FN$ represent true positives, true negatives, false positives, and false negatives, respectively, where the positive class corresponds to abnormal API requests and the negative class to normal API requests.

\section{Experimental Results}
\label{sec:results}

This section presents the experimental results evaluating the effectiveness of our proposed framework. We begin by analyzing the identification of reserved tokens during the input masking phase. Next, we assess our method using a custom-trained RoBERTa model, which we trained from scratch on the experimental datasets. We compare this model to the pre-trained RoBERTa-Base model to evaluate the impact of domain-specific language modeling. The evaluation involves measuring lexical similarity and semantic alignment with the original inputs to gauge the quality of the generated API requests. Additionally, we evaluate the impact of data augmentation on classification performance using a Random Forest model, before and after dataset extension. Furthermore, we apply our method to SOTA HTTP anomaly detection techniques to evaluate its potential for enhancing more advanced classifiers. Finally, to assess the sensitivity of our generative process to the reserved token filtering mechanism, we conducted an ablation study focused on the confidence level parameter used to derive the frequency threshold.

We used an Intel(R) Xeon(R) Silver 4214R CPU @ 2.40GHz to evaluate the effectiveness of each method. Each experiment was repeated five times, and we report the average score across all evaluation metrics. The custom-trained RoBERTa model was configured with 6 transformer layers, 12 attention heads, a hidden size of 768, and a vocabulary size of 52,000. The model was trained for 20 epochs with a batch size of 32 and a block size of 128 tokens. Warmup steps were set to 5\% of the total training steps. We used a maximum sequence length of 512 during tokenization and applied whole-word masking for MLM training. For the Random Forest classifier, the anomaly detection threshold was dynamically defined as the 99th percentile of the model output probabilities over the normal training data and used as a reference point for classification.

\subsection{Reserved Tokens}
To ensure that API functionality operates within a constrained structural language framework, we first identified a set of reserved tokens to exclude them from the generative process. This step helps preserve the syntactic and semantic integrity of generated API requests by preventing frequent structural tokens from being replaced. 
We used a z-score of 5.73, derived from the specified confidence level, to calculate a frequency threshold for each dataset. For CSIC 2010, the threshold was set at 6,177.41, identifying high-frequency tokens such as "=" and "\&". For ATRDF 2023, the threshold was 4,977.74, resulting in 32 reserved tokens including "/" and "get". Common structural tokens such as "/" and "=" appeared across both datasets, reflecting standard HTTP request patterns. Meanwhile, domain-specific tokens such as "tienda1" in CSIC 2010 and "check" in ATRDF 2023 highlighted differences between the two API environments.

\subsection{Baselines}

In this experiment, we produced one augmented sample for each input HTTP request in the training set. To ensure the quality of the synthetic data, only the generated samples that passed the discriminator's validation were retained. This process resulted in a validated synthetic version for every original request, effectively doubling the size of the training data in each scenario. Our objective in this evaluation is twofold: (1) to improve upon the GLASS-FODD~\cite{anidjar2024extending} framework in terms of the semantic and structural quality of the generated API requests, and (2) to assess whether higher-quality augmentation contributes to improved anomaly detection performance.

As shown in Table~\ref{tab:bleu}, the custom-trained RoBERTa generator achieved the highest scores across all three text similarity metrics on both datasets, with a few exceptions. On CSIC 2010, it outperformed GLASS-FODD by +5.74\% in BERTScore and +2.34\% in MoverScore, and slightly improved BLEU by +0.53\%. Compared to the RoBERTa-Base variant, the custom model showed a +0.76\% gain in BERTScore and +1.51\% in MoverScore, though BLEU was slightly lower by -0.80\%. The difference is more pronounced on ATRDF 2023, where the custom model achieved a BERTScore improvement of +11.10\% over GLASS-FODD and +2.21\% over RoBERTa-Base. It also led in MoverScore by +5.40\% and +4.35\%, respectively. However, in BLEU, RoBERTa-Base scored higher by +3.14\%.

This discrepancy can be attributed to the nature of the BLEU metric, which measures surface-level similarity by computing n-gram overlaps. Because our generation process replaces a single token in the original request with a semantically appropriate alternative, it often preserves meaning but not exact word sequences. The pretrained RoBERTa-Base model, having broader language coverage, may favor more common or expected token replacements that increase surface-level similarity, leading to slightly higher BLEU scores despite weaker semantic matching. In contrast, the custom model focuses on domain-specific terminology and structure, which leads to more accurate semantic replacements, as reflected in higher BERTScore and MoverScore values. BERTScore captures semantic similarity using contextual embeddings, while MoverScore uses semantic distance to better capture meaning despite lexical differences.

\renewcommand{\arraystretch}{1.15} % Slightly tighter vertical spacing
\setlength{\tabcolsep}{6pt}  
\begin{table*}[!ht]
    %\small
    \centering
    \footnotesize                 % or \scriptsize if space is tight
    \begin{tabular}{l|l|c|c|c}
    \hline
    \textbf{Dataset} & \textbf{GAN Architecture} & \textbf{BERTScore} & \textbf{MoverScore} & \textbf{BLEU} \\ \hline
    \multirow{3}{*}{\shortstack[l]{\textbf{CSIC}\\\textbf{2010}}}
        & Ours w/RoBERTa Custom & 96.74\% & 89.95\% & 95.27\% \\     
        & Ours w/RoBERTa-Base            & 95.98\% & 88.44\% & 96.07\% \\
        & GLASS-FODD                     & 91.00\% & 87.61\% & 94.74\% \\ \hline
    \multirow{3}{*}{\shortstack[l]{\textbf{ATRDF}\\\textbf{2023}}}
        & Ours w/RoBERTa Custom  & 93.04\% & 83.58\% & 86.94\% \\ 
        & Ours w/RoBERTa-Base             & 90.83\% & 79.23\% & 90.08\% \\
        & GLASS-FODD                      & 81.94\% & 78.18\% & 84.80\% \\ \hline
    \end{tabular}
    \caption{
    Semantic and structural quality evaluation of the proposed framework on the CSIC 2010 and ATRDF 2023 datasets, in comparison with the pretrained RoBERTa-Base language model and the other GAN-Inspired technique. BERTScore measures semantic similarity using contextual embeddings, MoverScore incorporates Earth Mover's Distance to quantify semantic closeness, and BLEU evaluates surface-level similarity based on n-gram overlap. Higher scores indicate stronger alignment between generated and reference API requests across different levels of textual fidelity.}
    
\label{tab:bleu}
\end{table*}

To evaluate how data augmentation affects classification performance, we used a Random Forest classifier trained twice for each configuration: once using only the original training data, and once using the original data combined with the generated samples. In both cases, the test dataset remains unchanged. The results shown in Table~\ref{tab:model-performance-1} include both the overall classification metrics and the absolute improvement observed after dataset extension. This setup allows us to isolate the contribution of the synthetic data to the final classification outcome.

On the CSIC 2010 dataset, the custom RoBERTa model achieves an MCC of 95.62\% and an F1 score of 97.59\%, with respective gains of +1.69\% and +1.00\% after augmentation. Although the improvements are relatively small, the baseline scores are already high, leaving a limited margin for further gain. In comparison, the RoBERTa-Base model shows a larger relative improvement in F1 score (+4.94\%), while GLASS-FODD sees only a modest increase (+0.25\%) despite starting from a lower baseline.

On the ATRDF 2023 dataset, the custom model reaches an MCC of 91.06\% (+19.53\%) and an F1 score of 91.24\% (+21.10\%), with the most substantial improvement observed in Precision (+38.82\%). Recall remains consistently high across all models at 99.97\%, indicating that augmentation mainly improves the classifier’s ability to reduce false positives rather than increase detection sensitivity.

A possible reason for GLASS-FODD's limited improvement lies in its design, which was originally intended for short SMS messages. Its generator selects tokens based on embedding dissimilarity but does not explicitly consider the semantic role or syntactic context of the replaced token, and any token in the vocabulary is a potential candidate for substitution. In the case of HTTP request logs, where structure, syntax, and token order are critical, such replacements can disrupt functional patterns and reduce the effectiveness of the generated samples. On the ATRDF 2023 dataset, which includes more diverse and natural language-like payloads written in English, GLASS-FODD performs slightly better, unlike CSIC 2010, which is primarily in Spanish and more syntactically strict.

\renewcommand{\arraystretch}{1.2}
\begin{table*}[!]
    %\small
    \centering
    %\resizebox{\textwidth}{!}{ % Resize to fit page width
    \begin{tabular}{p{1cm}|p{3.6cm}|p{2.7cm}|p{2.7cm}|p{2.7cm}|p{2.3cm}}
    \hline
    \textbf{Dataset} & \textbf{GAN Architecture} & \textbf{MCC} & \textbf{F1 Score} & \textbf{Precision} & \textbf{Recall} \\ \hline
    \multirow{3}{*}{\shortstack[l]{\textbf{CSIC}\\\textbf{2010}}}
        & Ours w/RoBERTa Custom & 95.62\% (+1.69\%) & 97.59\% (+1\%) & 95.79\% (+1.8\%) & 99.47\% (-0.03\%) \\  
        & Ours w/RoBERTa-Base   & 74.31\% (+11.36\%) & 86.17\% (+4.94\%) & 77.14\% (+9.95\%) & 97.6\% (-1.4\%) \\
        & GLASS-FODD            & 72.10\% (+0.99\%) & 85.03\% (+0.25\%) & 81.79\% (+6.15\%) & 88.54\% (-6.15\%)\\ 
    \hline
    \multirow{3}{*}{\shortstack[l]{\textbf{ATRDF}\\\textbf{2023}}}
        & Ours w/RoBERTa Custom & 91.06\% (+19.53\%) & 91.24\% (+21.1\%) & 83.92\% (+38.82\%) & 99.97\% (+0\%) \\ 
        & Ours w/RoBERTa-Base   & 93.67\% (+2.69\%) & 93.87\% (+2.69\%) & 88.45\% (+5.08\%) & 100\% (+0\%)\\
        & GLASS-FODD            & 88.92\% (+1.93\%) & 89.05\% (+2\%) & 80.29\% (+3.61\%) & 99.97\% (+0\%)\\ 
    \hline
    \end{tabular}
    %}
    \caption{Classification performance evaluated on the CSIC 2010 and ATRDF 2023 datasets using a Random Forest classifier. Values in parentheses indicate the absolute improvement over the corresponding model without dataset extension, highlighting the impact of synthetic data generation.}
    \label{tab:model-performance-1}
\end{table*}

\subsection{Anomaly Detection SOTA}

To assess the broader impact of our technique, we evaluated its effectiveness when applied to three SOTA HTTP anomaly detection methods: Sec2vec~\cite{gniewkowski2023sec2vec}, SWAF~\cite{jemal2022swaf}, and VLADEN~\cite{jagat2024detecting}. In this evaluation, we used only the custom-trained RoBERTa variant as the GAN architecture for generating augmented samples. The results are summarized in Table~\ref{tab:model-performance}.

Overall, the inclusion of generated samples led to improved performance across all methods and datasets. For Sec2vec, we observed consistent gains across all metrics, particularly in precision, with a smaller but stable increase in F1 score and MCC. This suggests that the augmented samples contribute useful variation that reinforces the model's ability to distinguish subtle patterns in request structure. SWAF also showed improved performance, especially in recall, while precision remained stable or showed slight variation. These results indicate that the augmented samples helped the model capture a broader set of anomaly patterns, with minor trade-offs in false positives on CSIC 2010, which is more constrained. The most notable improvement was observed in VLADEN, which initially performed significantly lower than the other models, particularly on ATRDF. After augmentation, VLADEN showed substantial gains on the ATRDF dataset, with an increase of +26.18\% in F1 score and +45.92\% in Precision, along with a +23.14\% increase in MCC. While improvements on CSIC 2010 were smaller, the upward trend confirms that exposure to syntactically consistent but semantically varied samples helped the model better reconstruct and identify anomalies.

\renewcommand{\arraystretch}{1.2}
\begin{table*}[!]
    %\small
    \centering
    \begin{tabular}{p{1.1cm}|p{1.5cm}|p{2.7cm}|p{2.7cm}|p{2.7cm}|p{2.65cm}}
    \hline
    \textbf{Dataset} & \textbf{Detector} & \textbf{MCC} & \textbf{F1 Score} & \textbf{Precision} & \textbf{Recall} \\ \hline
    \multirow{3}{*}{\shortstack[l]{\textbf{CSIC}\\\textbf{2010}}}
        & Sec2vec & 98.98\% (+1.55\%) & 99.04\% (+1.46\%) & 99.54\% (+2.56\%) & 98.54\% (+0.36\%)  \\  
        & SWAF    & 96.98\% (+2.15\%) & 98.33\% (+1.25\%) & 98.63\% (-0.78\%) & 98.04\% (+3.26\%) \\    
        & VLADEN  & 54.52\% (+3.85\%) & 76.59\% (+2.69\%) & 68.54\% (-4.61\%) & 86.78\% (+11.93\%)  \\   

    \hline
    \multirow{3}{*}{\shortstack[l]{\textbf{ATRDF}\\\textbf{2023}}}
        & Sec2vec & 97.8\% (+1.27\%) & 97.93\% (+1.2\%) & 98.94\% (+2\%) & 96.94\% (+0.31\%)  \\     
        & SWAF    & 99.35\% (+0.36\%) & 99.39\% (+0.34\%) & 98.99\% (+0.67\%) & 99.79\% (+0\%)  \\        
        & VLADEN  & 85.97\% (+23.14\%) & 85.97\% (+26.18\%) & 75.4\% (+45.92\%) & 100\% (+0\%)  \\   

    \hline
    \end{tabular}
    \caption{Performance comparison of the SOTA anomaly detection methods on the CSIC 2010 and ATRDF 2023 datasets using our proposed data augmentation technique. All results are reported using the custom-trained RoBERTa GAN-inspired variant. Values in parentheses represent the absolute improvement after dataset extension, highlighting the impact of data generation on detection performance.}
    \label{tab:model-performance}
\end{table*}

\subsection{Ablation Study}

In the generative process, building an effective token vocabulary involves setting an appropriate confidence level to derive the frequency threshold for reserved tokens. As described in Section~\ref{sec:input_masking}, this threshold is used to identify high-frequency structural tokens, such as delimiters and protocol-specific keywords that are excluded from masking and replacement to preserve the syntactic and semantic integrity of generated API requests. 

To assess how sensitive our approach is to the choice of confidence level, we conducted an ablation study that systematically varied this parameter. Specifically, we evaluated the impact of different confidence levels on detection performance by adjusting the threshold for reserved token filtering. The study was performed using the custom-trained RoBERTa model. In our default setup, we applied a conservative confidence level of 99.99\%. For the ablation, we varied the confidence level from 97\% to 99.5\% and measured the resulting F1 scores. 

\begin{figure}[!htb]
\centering
\begin{subfigure}{0.4\textwidth}
    \centering
    \includegraphics[width=\linewidth]{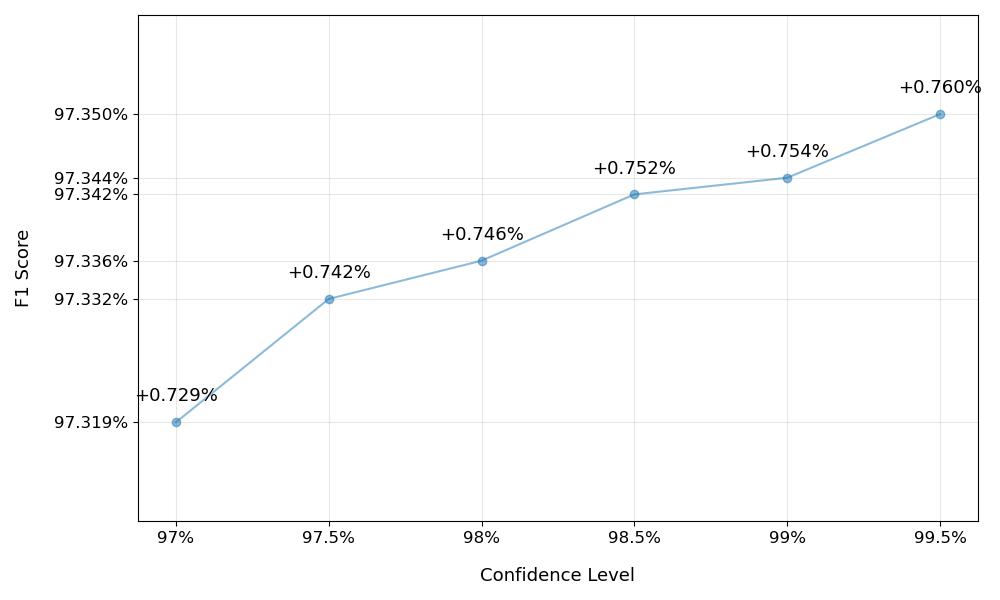}
    \caption{CSIC 2010}
    \label{fig:csic_2010}
\end{subfigure}
\hfill
\begin{subfigure}{0.4\textwidth}
    \centering
    \includegraphics[width=\linewidth]{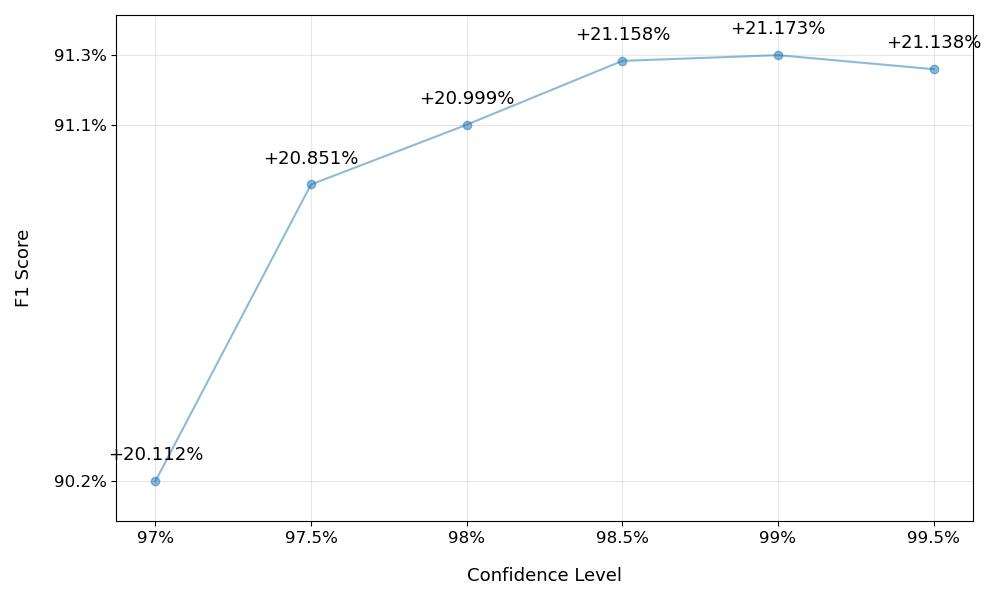}
    \caption{ATRDF 2023}
    \label{fig:atrdf_2023}
\end{subfigure}
\caption{Impact of varying the confidence level used to compute the reserved token frequency threshold on classification performance. The plots show the F1 score improvement (after dataset extension) across different thresholds for the CSIC 2010 and ATRDF 2023 datasets. Higher confidence levels exclude more frequent structural tokens from replacement during generation.}
\end{figure}

As shown in Figure~\ref{fig:csic_2010}, CSIC 2010 shows modest and steady improvements, with F1 score gains rising from +0.729\% at 97\% to +0.760\% at 99.5\%, and reaching +1.00\% at our default confidence level of 99.99\%~\ref{tab:model-performance}. Although these differences are modest in absolute terms, even minor gains can translate into significant impact at scale, where hundreds of thousands of API requests may be classified daily. In contrast, as can be seen in Figure~\ref{fig:atrdf_2023}, the ATRDF 2023 dataset shows a peak improvement of +21.173\% at 99\%, followed by a slight decline to +21.138\% at 99.5\% and +21.10\% at 99.99\%~\ref{tab:model-performance}. These results suggest that further improvements can be achieved by selecting an appropriate confidence level for the specific use case, while also demonstrating that our approach remains stable and effective across different settings.

As part of our evaluation, we also compared the data generation times between the pretrained RoBERTa-Base model and a custom-trained RoBERTa model. The RoBERTa-Base model was trained on over 160 GB of diverse English-language corpora, which included web content, news articles, books, and encyclopedic texts~\cite{liu2019roberta}. In contrast, the custom model was trained exclusively on experimental datasets, utilizing 3.3 MB of HTTP traffic for CSIC 2010 and 4.4 MB for ATRDF 2023. Given the significantly smaller training data, the custom-trained model exhibited faster generation times, averaging 0.526 seconds per input sample, compared to 0.957 seconds for RoBERTa-Base.

\section{Discussion and Conclusions}
\label{sec:conclusions}

In this paper, we presented a novel GAN-inspired learning framework for enhancing API anomaly detection through domain-aware data augmentation. By integrating a custom-trained RoBERTa model into both the generation and discriminator processes, our method synthesizes semantically meaningful API requests that preserve structural integrity while introducing useful variability. Through a targeted masking strategy guided by semantic distance and frequency-based token filtering, we ensured that the generated requests align closely with real-world API traffic patterns.

Our experimental results across two benchmark datasets, CSIC 2010 and ATRDF 2023, demonstrate that the proposed framework improves both the semantic quality of synthetic samples and the effectiveness of downstream anomaly detection models. Compared to a previous GAN-inspired technique, our method achieved stronger semantic alignment, and led to measurable gains in classification performance using Random Forests. Furthermore, when integrated into three SOTA anomaly detection architectures, our method consistently boosted precision and F1 scores, with the most significant gains observed in VLADEN on ATRDF (+26.18\% F1 and +45.92\% Precision).

An ablation study revealed that our approach is robust to changes in the confidence level used for token filtering, although small adjustments to this parameter can yield additional performance gains, particularly in datasets with diverse or loosely structured API inputs.

We also observed that our custom-trained RoBERTa model generated high-quality augmentations more efficiently than the larger RoBERTa-Base model, despite being trained on significantly smaller, domain-specific datasets. This highlights the value of compact, task-specific language models.

Currently, the framework generates a single validated synthetic request per input sample. Future work should explore generating multiple diverse augmentations per request to improve coverage, especially for less common attack patterns. In addition, this study did not explicitly balance attack categories during training, future research should investigate class-aware augmentation strategies to better support rare or emerging threats.

\section*{Funding}
This work was supported by the Israel Innovation Authority and the Trust-AI consortium.

\section*{Acknowledgment}
This work was supported by the Ariel Cyber Innovation Center in conjunction with the Israel Innovation Authority, and the Trust-AI consortium. 

\printcredits

\bibliographystyle{cas-model2-names}

\bibliography{cas-refs}

\newpage
\vskip3pt

\bio{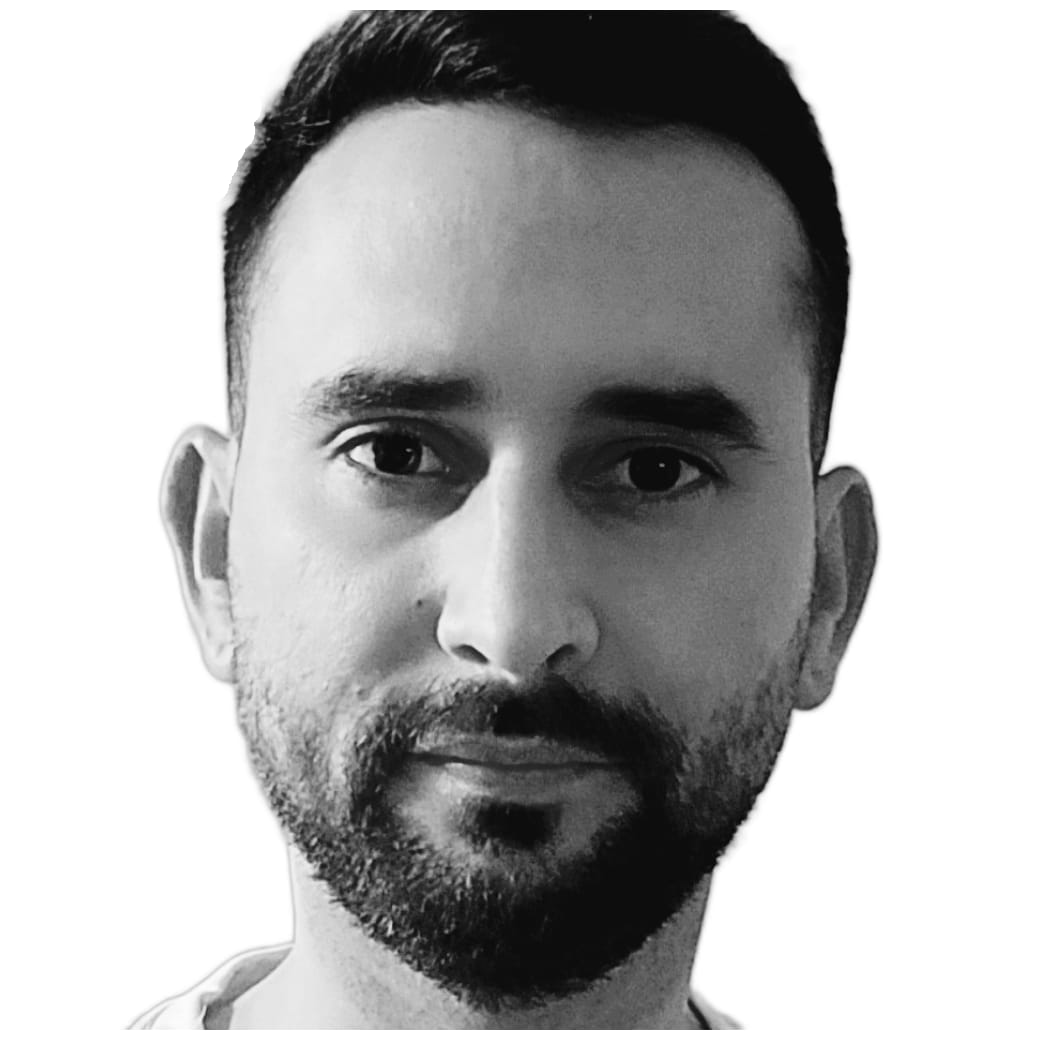}
Udi Aharon is currently pursuing a Ph.D. degree in the Department of Computer and Software Engineering at Ariel University, Israel. His research activities span the fields of Machine Learning and Cybersecurity. Specifically, his work focuses on enhancing API security through the application of text-based models.
\endbio

\bio{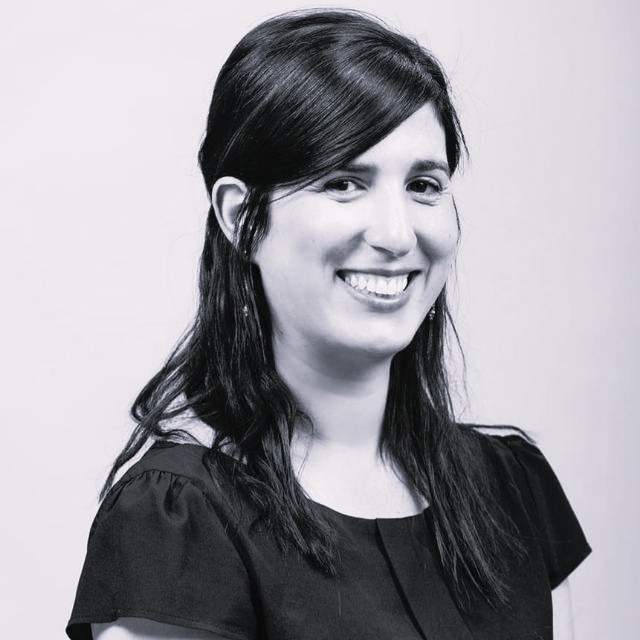}
Dr. Revital Marbel received her B.Sc., M.Sc., and Ph.D. degrees in Computer Science from Ariel University.
Today, Revital Marbel is a faculty member at Holon Institute of Technology (HIT).
Her main research  include the use of NLP Generative Adversarial Network (GAN) and graph neural network (GNN) models for cybersecurity issues like SMS frauds and API and cloud security.
\endbio

\bio{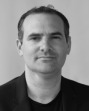}
Dr. Ran Dubin received his B.Sc., M.Sc., and Ph.D. degrees from Ben-Gurion University, Beer Sheva, Israel, all in communication systems engineering. He is a faculty member at the Computer and Software Engineering Department, Ariel University, Israel. His research interests revolve around zero-trust cyber protection, malware disarms and reconstruction, encrypted network traffic detection, Deep Packet Inspection (DPI), bypassing AI, Natural Language Processing, and AI trust.
\endbio

\bio{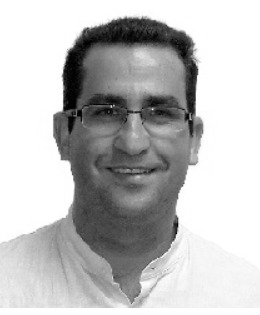}
Prof. Amit Dvir received his B.Sc., M.Sc., and Ph.D. degrees in communication systems engineering from Ben-Gurion University, Beer Sheva, Israel. He is currently the head of the Computer and Software Engineering Department and the head of the Ariel Cyber Innovation Center at Ariel University, Israel. From 2011 to 2012, he was a Postdoctoral Fellow at the Laboratory of Cryptography and System Security, Budapest, Hungary. His research interests include enrichment data from encrypted traffic.
\endbio

\bio{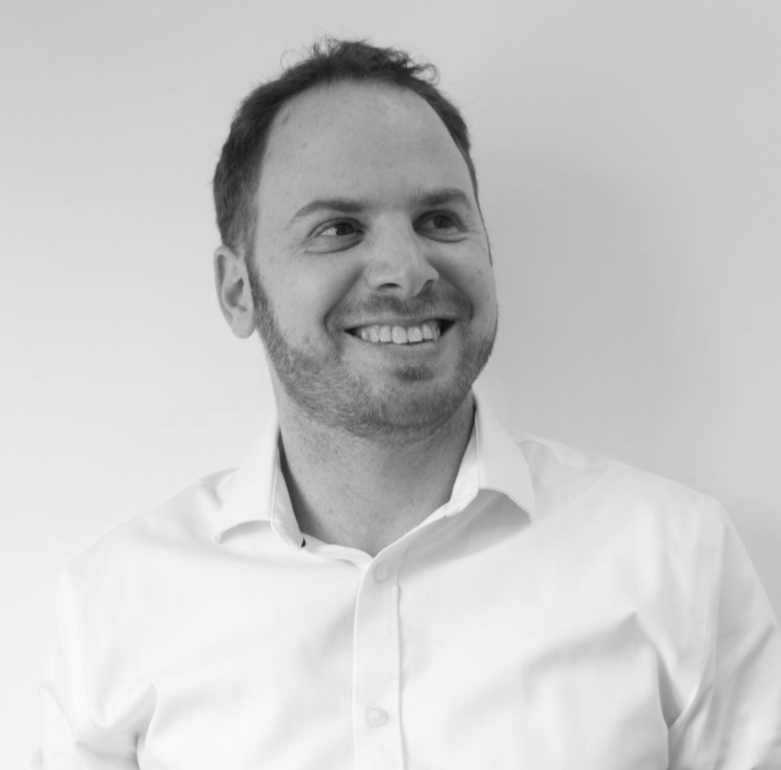}
Dr. Chen Hajaj holds a Ph.D. in Computer Science, an M.Sc. in Electrical Engineering, and a B.Sc. in Computer Engineering, all from Bar-Ilan University. He is a faculty member in the Department of Industrial Engineering and Management and a researcher at the Ariel Cyber Innovation Center. He completed his postdoctoral fellowship at Vanderbilt University (2016–2018). His research focuses on machine learning, game theory, and cybersecurity, with expertise in encrypted traffic classification, adversarial AI, and multimodal learning.
\endbio

\end{document}